\documentclass[a4paper]{article}

\usepackage{romp32}
\usepackage{amsfonts,latexsym%
  ,amsmath,bbm%
  }
\newcommand{\PCPcontraction}{%
  \mathbin{%
    \hbox{%
      \!\vrule height 0.4pt width 5.0pt%
      }%
    \hbox{\vrule height 6.0pt width 0.4pt}\;
    }%
}%
\newcommand{\PCPoV}{\ensuremath{{\Omega_{\Gamma}}}}
\newcommand{\PCPtV}{\ensuremath{{\Theta_{\Gamma}}}}
\newcommand{\PCPdV}{\ensuremath{{d_{\bar\Gamma}}}}
\def\PCPqed{\hspace*{\fill}\ensuremath{\Box}\\}
\begin{document}

\title{
On The Geometry of Field Theoretic Gerstenhaber Structures%
}

\author{Cornelius Pauf\/ler\\
Fakult\"at f\"ur Physik\\
Albert-Ludwigs-Universit\"at Freiburg im Breisgau\\
Hermann-Herder-Stra\ss e 3\\
D 79104 Freiburg i. Br.\\
e-mail: pauf\/ler@physik.uni-freiburg.de
}
\date{July 21, 2000}

\maketitle

\begin{abstract}
Field theoretical models with first order Lagrangean can be formulated
in a covariant Hamiltonian formalism. In this article, the geometrical
construction of the Gerstenhaber structure that encodes the equations
of motion is explained for arbitrary fibre bundles. 
Special emphasis has been put on naturality of the constructions.
Further, the treatment of symmetries is explained. Finally, the 
canonical field theoretical 2-form is obtained by pull back and integration
of the polysymplectic form over space like hypersurfaces.
\end{abstract}

\noindent
{\bf Key words:} Geometric field theory, Multisymplectic geometry, Hamiltonian
formulation, Gerstenhaber algebra
\section{Introduction}
It has turned out that field theories
can be described best in the jet bundle formalism. For first order
theories, i.e. when the Lagrange density is a function of the fields
and their first derivatives only, the description can be reduced to
the first jet bundle. 
This description 
is an extension of the Lagrangean formulation of mechanics in the
following sense. In particle mechanics,
particles are described by curves in the tangent space $T\mathcal Q$, i.e. by
sections of the (trivial) bundle $\mathbbm R\times T\mathcal Q$.
In field theory, physical fields are sections of some bundle over
space-time, i.e. higher dimensional curves. Consequently, the base
manifold of a given field theory -- which quite often is space-time --
{\em should not be confused with the configuration space of
  mechanics.} Rather, the latter is represented by the fibres of the
given fibre bundle when the base manifold is $\mathbbm R$, i.e. the
time axis.\\
The corresponding field equations are second order but -- for nicely
behaving Lagrange densities -- can be 
reduced to a system of first order equations, the so-called de
Donder--Weyl equations (for a review see \cite{GoIsMa98}). One might
view these equations as a generalisation of the usual Hamilton
equations in mechanics. However, a peculiar feature of this approach is that 
there are as many conjugate momenta corresponding to each field coordinate
degree of freedom as there are space-time dimensions.\\
For the study of quantum effects in field theory, at least in the
sense of star products (\cite{QFT-Sterne}) there is the need
of an algebraic formulation of the theory. There has been recent
progress on that, namely the Gerstenhaber structure found by
Kanatchikov (\cite{Ka97,Ka98}).\\  
In this article, a geometric construction of this Gerstenhaber
structure will be proposed that works for arbitrary fibre bundles.
Particular attention will be paid to the treatment of internal
symmetries and to the correspondence to the standard Hamiltonian
formulation of field theory.

\section{\label{jet-notation}Phase space for first order field theories}

Let $(\mathcal E,\pi,\mathcal M)$ be a fibre bundle with $N$
dimensional fibres over an $n$ dimensional orientable 
manifold $\mathcal M$. Its first jet
bundle, $\mathfrak J^1\mathcal E$ (cf. \cite{Sa89}), consists of
equivalence classes of local sections of $\mathcal E$: Two sections
$\varphi$ and $\psi$ are considered equivalent at a point 
$m\in \mathcal M$, if they take the same value in $m$, 
$\varphi(m)=\psi(m)$,  and their tangent mappings coincide,
$T_m\varphi=T_m\psi$. The class defined by a section $\varphi$ 
is denoted by $j^1_m\varphi$.
$(\mathfrak J^1\mathcal E,\pi_1,\mathcal M)$ is a fibre bundle of fibre
dimension $(Nn+N+n)$, while $(\mathfrak J^1\mathcal E,\pi^1_0,\mathcal E)$
is an affine bundle with associated vector bundle 
$(\mathfrak V\mathcal E\otimes T^\ast\mathcal M,
  \tau_{\mathcal E}\otimes\tau^\ast_{\mathcal M},\mathcal E)$ (cf. \cite{Sa89}).
Here, $\mathfrak V\mathcal E$ denotes the vertical (w.r.t. $\pi$) 
tangent bundle on $\mathcal E$, while $\tau_{\mathcal E}$ and 
$\tau^\ast_{\mathcal M}$ stand for the
respective canonical projections. Note that $T^\ast\mathcal M$ is in
fact pulled back onto $\mathcal E$ -- this has been suppressed for simplicity.  

Let the dual jet bundle
$\left((\mathfrak J^1\mathcal E)^\star,
  (\pi^1_0)^\star,\mathcal E\right)$
be the fibre bundle over $\mathcal E$ with fibres the affine, fibre
preserving mappings from $\mathfrak J^1\mathcal E$ to 
$\Lambda^nT^\ast\mathcal M$, i.e. for every $v\in\mathcal E$ one has
\begin{equation*}
  \left((\mathfrak J^1\mathcal E)^\star\right)_v=
  \{A:(\mathfrak J^1\mathcal E)_v
  \rightarrow \Lambda^n(T^\ast\mathcal M)_{\pi(v)} \; \textrm{affine}\}.
\end{equation*}
$(\mathfrak J^1\mathcal E)^\star$ inherits a vector space structure
from $\Lambda^n T^\ast\mathcal M$. 

As a connection provides a map 
$\Gamma:\mathcal E\rightarrow\mathfrak J^1\mathcal E$ (see appendix
\ref{conn-facts} for details), there is an
isomorphism
\begin{equation}\label{def-i-Gamma}
  i_\Gamma:\mathfrak J^1\mathcal E 
  \stackrel\cong\rightarrow \mathfrak V\mathcal E\otimes T^\ast\mathcal M
\end{equation}
that sends the image of $\Gamma$ to the zero section. This induces a
splitting of the dual jet bundle 
$\left((\mathfrak J^1\mathcal E)^\star\right)$,
\begin{equation}\label{i-Gamma-star}
  (i_\Gamma)^\star:
  (\mathfrak J^1\mathcal E)^\star
  \stackrel\cong\rightarrow
  \left((\mathfrak V\mathcal E)^\ast\otimes T\mathcal M\right)
  \otimes \Lambda^nT^\ast\mathcal M
  \oplus \Lambda^nT^\ast\mathcal M =:\mathcal P\oplus\mathcal L,
\end{equation}
where we have introduced the notation $\mathcal P$ for the first
summand and $\mathcal L$ for the latter. Observe that $\mathcal L$ is
a line bundle over $\mathcal E$ which is trivial for orientable
manifolds. 
$\mathcal P$ will be called the 
{\em polysymplectic phase space} from now on. 

\subsection{Coordinates}

Let $(x^\mu,v^A)$\footnote{
For the rest of this article, lower case Greek indices
$\mu,\nu,\rho,\sigma,\ldots$ label directions on
the base manifold $\mathcal M$, i.e. run from $1$ to $n$, while
capital Latin characters $A,B,C,D,\ldots$ are used for 
fibre coordinates of $\mathcal E$, i.e. run from $1$ to $N$.
}
be local
coordinates of $\mathcal E$. Then to every point $j^1_m\varphi$ 
in $\mathfrak J^1\mathcal E$ one can assign the matrix elements of 
$T_m\varphi$ in those coordinates, hence we have locally a set 
$(x^\mu,v^A,v^A_\mu)$ for each such point. Further, the choice of
coordinates in $\mathcal E$ induces a trivialisation of the dual jet
bundle: Every affine map $\underline p$ can be written as
$
  \underline p:(x^\mu,v^A,v^A_\mu)\mapsto (x^\mu,v^A,(p^\mu_A v^A_\mu+p)d^nx),
$ hence $(x^\mu,v^A,p^\mu_A,p)$ provides a set of local coordinates.
Then, using the isomorphism $(i_\Gamma)^\star$, (\ref{i-Gamma-star}),
one can show (\cite{Pa00}) that the first three sets of coordinates can
$(x^\mu,v^A,p^\mu_A)$ can be used to label 
$\mathcal P$ locally. Furthermore, the arbitrariness of the choice of a
connection connection $\Gamma$ is projected out, hence the projection
$\pi_{\mathcal P}$ from $(\mathfrak J^1\mathcal E)^\star$ onto
$\mathcal P$ is canonical.

\subsection{$\bar\Gamma$-Vertical differentials}

\begin{definition}{Definition}[$\bar\Gamma$-vertical differential]\\
  Let $P_{\bar\Gamma}$ be the projection of $T\mathcal P$ onto 
  its vertical subbundle $\mathfrak V\mathcal P$ constructed from a connection 
  $\bar\Gamma:\mathcal P\rightarrow\mathfrak J^1\mathcal P$, 
  cf. (\ref{conn-facts}).
  The corresponding $\bar\Gamma$-vertical differential $\PCPdV$ is the map
  \begin{equation*}
    \PCPdV:\Omega^r(\mathcal P)\rightarrow\Omega^{r+1}(\mathcal P)
    \quad (\forall r)
  \end{equation*}
  satisfying the following properties ($d$ denotes the exterior
  derivative on $\mathcal P$)
  \begin{enumerate}
  \item For $f\in\mathcal C^\infty(\mathcal P)$, 
    $X\in\mathfrak X(\mathcal P)$, one has
    $
      (\PCPdV f)(X)=(df)(P_{\bar\Gamma}X).
    $
  \item $\PCPdV$ acts as a graded derivation of degree $1$. 
    For $\alpha,\beta\in\Omega^\ast(\mathcal P)$,
    \begin{equation*}
      \PCPdV(\alpha\wedge\beta)
      =(\PCPdV\alpha)\wedge\beta
      +(-)^{|\alpha|}\alpha\wedge(\PCPdV\beta).
    \end{equation*}
  \item $(\PCPdV)^2=0$.
  \item $\PCPdV$ vanishes on pulled back forms, i.e. for  
    $\alpha\in\Omega(\mathcal M)$, 
    $
      \PCPdV \left((\pi_1)^\star\right)^\ast\alpha=0.
    $
  \end{enumerate}
\end{definition}
One can indeed show that for every $\bar\Gamma$ there is a unique
$\PCPdV$. 
Note that $d$ and $\PCPdV$ do not anticommute.\\
In what follows, we fix the connection to be the map
$\bar\Gamma$ defined in the appendix, (\ref{Gamma-bar-coord}).
Then $\PCPdV$ is
locally given by
\begin{equation}
  \PCPdV f
  =\partial_A f\mathfrak e^A+\partial_\mu^A f \mathfrak e^\mu_A,
  \quad
  \PCPdV dx^\mu=0=\PCPdV\mathfrak e^A=\PCPdV\mathfrak e^\mu_A,
\end{equation}
where $f\in\mathcal C^\infty(\mathcal P)$ 
is a function and the forms $\mathfrak e^A$ and $\mathfrak e^\mu_A$ are defined by
(\ref{Gamma-basis}) and  together with the $dx^\mu$ form a basis of 
$T^\ast\mathcal P$.

\subsection{Canonical forms}

As 
$\mathcal P=\left((\mathfrak V\mathcal E)^\ast\otimes T\mathcal M)
  \otimes\Lambda^nT^\ast\mathcal M\right)$ and fibres of
$\Lambda^nT^\ast\mathcal M$ are one dimensional, every point 
of $\mathcal P$ can be written as $\alpha\otimes\omega$, where
$\alpha$ and $\omega$ are elements of
$(\mathfrak V\mathcal E)^\ast\otimes T\mathcal M$
and $\Lambda^nT^\ast\mathcal M$ respectively. 
$\alpha$ can be viewed as a map 
$\mathfrak V\mathcal E \rightarrow T\mathcal M$.
Define an $n$-form $\PCPtV$ on $\mathcal P$ by
\begin{multline}
  {\left(\PCPtV\right)}_{(\alpha\otimes\omega)}\left(X_1,\ldots,X_n\right)\\
  =
  \sum_\sigma {\scriptstyle\frac{(-)^\sigma}{n!}}
  \omega\left(\alpha(P_\Gamma\circ T(\pi^1_0)^\star(X_{\sigma(1)})),
    T(\pi_1)^\star X_{\sigma(2)},\ldots,T(\pi_1)^\star X_{\sigma(n)}
  \right).
\end{multline}
As $\omega$ is linear in each argument, the right hand side is well
defined, i.e. does not depend on the splitting $\alpha\otimes\omega$.
$\PCPtV$ generalises the construction of the canonical $1$-form on
a cotangent bundle, but due to the fact that $\mathcal E$ is
non trivial one has to choose a connection. 

The {\em polysymplectic 
$n+1$-form} is defined through the action of $\PCPdV$ on $\PCPtV$,
\begin{equation}
  \PCPoV=-\PCPdV\PCPtV.
\end{equation}
Using coordinates as above, one finds
\begin{equation}
  \begin{split}
    \PCPtV_{(x,v,p)}
    &={\scriptstyle \frac1{(n-1)!}}
    \epsilon_{\mu\nu_1\cdots\nu_{n-1}}
    p^\mu_A dv^A\wedge dx^{\nu_1}\wedge\cdots\wedge
    dx^{\mu_{n-1}}
    +p^\mu_A\Gamma^A_\mu d^nx,
    \\
    \PCPoV_{(x,v,p)}
    &={\scriptstyle \frac1{(n-1)!}}\epsilon_{\mu\nu_1\cdots\nu_{n-1}}
    dv^A\wedge dp^\mu_A\wedge
    dx^{\nu_1}\wedge\cdots\wedge dx^{\mu_{n-1}}
    - \PCPdV (p^\mu_A\Gamma^A_\mu d^nx).
  \end{split}
\end{equation}
Note that we have used
(\ref{Gamma-spur}) and that $\PCPoV$ indeed does depend on $\Gamma$
rather than on $\bar\Gamma$. Moreover, the $\PCPdV$-cohomology class 
of \PCPoV\ is independent of $\Gamma$.

\section{Covariant Hamiltonian dynamics}

\begin{definition}{Definition}[Hamiltonian forms]\\
  Let $\PCPdV$ be the vertical differential to $\bar\Gamma$ of (\ref{Gamma-bar-coord}). A
  horizontal form $H\in\Omega(\mathcal P)$ is called Hamiltonian 
  if there exists a multi-vector $X_H$, i.e. 
  an element in $\Lambda\mathfrak X(\mathcal P)$, that satisfies
  \begin{equation}\label{Ham-Vf}
    \PCPdV H = X_H \PCPcontraction \PCPoV.
  \end{equation}
  In this case, $X$ is called Hamiltonian multi-vector field.
  The set of all Hamiltonian forms is denoted by $\mathcal H$.
\end{definition}
Note that not every horizontal form is Hamiltonian. Rather, 
condition (\ref{Ham-Vf}) imposes a severe restriction on such forms.
\begin{theorem}{Proposition}\label{momentum-dep}
    [Multimomentum dependence of Hamiltonian
  forms, \cite{Ka97}]\\
  Let $H\in\mathcal H$ be a Hamiltonian form of degree $|H|$. If 
  $|H|>0$, then, using coordinates as above, 
  \begin{equation}
    H_{(x,v,p)}=
    \sum_{r=0}^{n-|H|}{\scriptstyle\frac1{(|H|)!}}
    p^{\mu_1}_{A_1}\cdots p^{\mu_r}_{A_r} 
    h^{A_1\cdots A_r\mu_{r+1}\cdots\mu_{n-|H|}}_{(x,v)} 
    \epsilon_{\mu_1\cdots\mu_n}
    dx^{\mu_{n+1-|H|}}\cdots dx^{\mu_n},
  \end{equation}
  where the functions $h^{A_1\cdots A_r\mu_{r+1}\cdots\mu_{n-|H|}}$
  are pulled back from  $\mathcal E$.
\end{theorem}
For a proof, we refer to \cite{Ka97,Pa00}. There is no restriction on
functions on $\mathcal P$. Section \ref{Symmetries} provides
(standard) 
examples for Hamiltonian $(n-1)$-forms.

There is an additional operation on $(\pi_1)^\star$-horizontal forms, namely
the lifted Hodge-$\ast$ operation on $\mathcal M$. Of course, $\ast$ is invertible.

\begin{definition}{Definition and theorem}[Gerstenhaber structure,
  \cite{Ka97,Ka98}]\\
  The maps 
  \begin{equation}
    \begin{split}
      \{,\}:\mathcal H\times\mathcal H\ni(F,G)
      &\mapsto
      \{F,G\}=(-)^{n-|F|}X_F\PCPcontraction X_G\PCPcontraction\PCPoV
      \in\mathcal H,\\
      \bullet:
      \mathcal H\times\mathcal H\ni(F,G)
      &\mapsto F\bullet G=\ast^{-1}\left(\ast F\wedge\ast G\right)
      \in\mathcal H
    \end{split}
  \end{equation}
  are well defined (for the definition of $\ast$ see the remark above)
  and satisfy (let $F,G,H\in\mathcal H$ be horizontal forms 
  of (exterior) degree $|F|$, $|G|$, and $|H|$, respectively)
  \begin{enumerate}
  \item $\{F,G\}=-(-)^{(n+1-|F|)(n+1-|G|)}\{G,F\}$,
  \item $F\bullet G=(-)^{(n-|F|)(n-|G|)}G\bullet F$,
  \item $(-)^{(n+1-|F|)(n+1-|H|)}\{F,\{G,H\}\}
            +(-)^{(n+1-|G|)(n+1-|F|)}\{G,\{H,F\}\}$\\
        \hspace*{40ex}
        $+(-)^{(n+1-|H|)(n+1-|G|)}\{H,\{F,G\}\}=0$,
  \item $\{F,G\bullet H\}=\{F,G\}\bullet H
            +(-)^{(n+1-|F|)(n-|G|)}G\bullet\{F,H\}$.  
  \end{enumerate}
  Further, the bracket does not depend on the specific choice of $\bar\Gamma$.
\end{definition}
A detailed proof can be found in \cite{Pa00}. Note that, in
particular, $\{F,G\}$ does not depend on the ambiguity in the
correspondence of Hamiltonians $F$ and $G$ and their 
Hamiltonian multi-vector fields $X_F$, $X_G$.

\section{Legendre transformation and canonical 2-form on the space of solutions}

In this section, we try to establish a link between the
multisymplectic formalism and the standard Hamiltonian formulation of
field theory. 
The Euler--Lagrange equations of a given field theory are partial
differential equations. Typically a well posed initial value problem
for such a set of field equations consists of the specification of two
sets of functions
\begin{equation}\label{initial-data}
  \left(\varphi^A(x),\pi_A(x)\right)_{A=1,\ldots,N}
\end{equation}
on some hypersurface $\Sigma$ of $\mathcal M$. $\pi_A$ is known as the
momentum variable to the field $\varphi_A$. The corresponding solution
is a set of functions $\{\Phi^A(x,t)\}$ ($t$ labels the additional
coordinate in $\mathcal M$) on $\mathcal M$. The initial data
(\ref{initial-data}) span the field theoretical phase space 
$\mathcal P_{F.T.}$, while the fields $\Phi$ span the covariant
section space $\mathcal P_{cov}$ (the name
has been chosen to stress the analogy to the path space of classical
mechanics,
cf \cite{AbMa}) which typically is some subspace of 
$\Gamma(\mathcal E)$. 
Despite the similarity in the names, $\mathcal P_{F.T.}$ should not be
confused with the polysymplectic phase space, $\mathcal P$. In
particular, both the field theoretical phase space and the section
space are infinite dimensional spaces, while $\mathcal P$ is of
dimension $(n+1)(N+1)-1$. The Legendre map provides a map
from the section space to $\mathcal P_{F.T.}$, which is explicitely given
by
\begin{equation}
  \label{Legendre}
  \varphi^A(x)=\Phi^A(x,0),
  \quad \pi_A(x)
  =\frac{\partial L}{\partial \partial_t\Phi^A}
  \left((x,0),\Phi(x,0),\Phi'(x,0)\right),
\end{equation}
where $L$ is the Lagrange density and $\Phi'$ denotes all derivatives
of the fields $\Phi$. On the field theoretical phase space there is a
canonical (weak symplectic) 
$2$-form that can be pulled back using the tangent map of
(\ref{Legendre}). Thus there is a (non-canonical) $2$-form on
$\omega_L$ on the section space $\mathcal P_{cov}$. The vector fields
which can be plugged into $\omega_L$ are infinitesimal (vertical)
variations of the fields $\Phi$. They form the tangent space of
$\mathcal P_{cov}$ (we leave problems arising from the infinite
dimensionality aside), which therefore consists of all sections of the
vertical bundle $\mathfrak V\mathcal E$ to $\mathcal E$. The explicit
expression for the contraction of $\omega_L$ with two such vectors
$X$, $Y$ is found to be (cf. \cite{Wo92} for details)
\begin{equation}\label{def-omega-L}
  \omega_L(X,Y)={\textstyle\frac12}
  \int_\Sigma
  \left(
    {\textstyle\frac{\partial^2L}{\partial\varphi^A\partial\varphi^B_\nu}}
    (X^AY^B-Y^AX^B) 
    +
    {\textstyle\frac{\partial^2L}{\partial\varphi^A_\mu\partial\varphi^B_\nu}}
    (Y^B\nabla_\mu X^A-X^B\nabla_\mu Y^A)
  \right)
  d_\nu x,
\end{equation}
where $\nabla$ denotes a covariant derivative on sections of $\mathcal
E$ (and hence on those of  $\mathfrak V\mathcal E$). The
$\nu$-summation of this formula reduces to the $t$-component if the
normal direction of $\Sigma$ happens to be the time coordinate.
On the other hand, let $\tilde X$ be the prolongation to the vertical bundle
$\mathfrak V\mathcal E$ of a given variational field $X$. Explicitely, one finds
\begin{equation}
  \label{prol-variation}
  \tilde X^A(x,t)=\left(X^A,\nabla_\mu X^A(x,t)\right),
\end{equation}
where we have used the Christoffel symbols of $\nabla$ for the map
$i_\Gamma$, (\ref{def-i-Gamma}). Now one can use the Legendre
transformation and the connection $\Gamma$ to pull back the
polysymplectic $(n+1)$-form onto $\mathfrak V\mathcal E$. The
contraction of the resulting $(n+1)$-form with two prolonged
variational fields $\tilde X$, $\tilde Y$ yields a horizontal
$(n-1)$-form, which can be integrated over $\Sigma$. This yields
back the expression (\ref{def-omega-L}). Indeed, a calculation shows
\begin{multline*}
  {\textstyle
    \left((\mathfrak F\mathcal L)^\ast\PCPoV\right)_{(x,v,v')}
    =
    (\partial_B\partial_A^\mu L)dv^A\wedge dv^B\wedge d_\mu x
    +(\partial_B^\nu\partial_A^\mu L)
    dv^A \wedge dv^B_\nu \wedge d_\mu x
    }\\
  {\textstyle
    -\big(\Gamma_\mu^A (\partial_B\partial_A^\mu L)dv^B
    +\Gamma_\mu^A(\partial_B^\nu\partial_A^\mu L)dv^B_\nu
    +(\partial_A^\mu L)\partial_B\Gamma^A_\mu
    -(\partial_\mu\partial^\mu_A L) dv^A\big)\wedge dx
   },
\end{multline*}
where $(\mathfrak F\mathcal L)$ denotes the Legendre map and 
the repeated use of the map $i_\Gamma$ in the pull back has been
suppressed in the notation. Note that the last line vanishes on two
vertical vectors which establishes the following result.
\begin{theorem}{Lemma}\label{symplectic-2-form}
  Let $X,Y$ be field variations on $\mathcal M$, i.e. $\pi$-vertical
  tangent vector fields on $\mathcal E$. Let $\tilde X,\tilde Y$ be
  their lifts via the Legendre mapping (\ref{Legendre}) --- taken for
  arbitrary value of $t$ --- to $\mathcal P$. Then on every section
  $\Phi\in\Gamma(\mathcal E)$ we find 
  \begin{equation}
    \omega_L(X\circ\Phi,Y\circ\Phi)
    =\int_\Sigma\;
    {\tilde\Phi}^\ast
    \Big(\tilde Y\PCPcontraction (\tilde X\PCPcontraction \Omega_\Gamma)\Big),
  \end{equation}
  where $\tilde\Phi$ denotes the prolongation of $\Phi$ to $\mathcal P$
  using (\ref{Legendre}) again.
\end{theorem}
The lemma shows how the standard Hamiltonian formulation of field
theory and the polysymplectic approach can be related. Objects (vector
fields in this case) in
polysymplectic fields theory have to be evaluated on (prolonged)
sections and
integrated over the submanifold $\Sigma$ that has been chosen for the
Hamiltonian picture. For instance, the variations $X$ and $Y$ are
given for all sections, but when evaluated on a specific $\Phi$, they
become variations of that $\Phi$.

\section{\label{Symmetries}Internal symmetries}

From the peculiar momentum dependence of multisymplectic Hamiltonian
$(n-1)$-forms as stated in proposition 3 
one immediately concludes the following.
\begin{theorem}{Lemma}
  Let $X$ be a Hamiltonian $1$-vector field in the sense of
  (\ref{Ham-Vf}). 
  Then $X$ is projectable
  along $(\pi^1_0)^\star$ onto $\mathcal E$. Conversely, the canonical
  lift $\hat X$ to $\mathcal P$ of a $\pi$-vertical
  vector field $X$ on $\mathcal E$ is Hamiltonian. Its Hamiltonian
  form is given by $I(X)=\hat X\PCPcontraction \PCPtV$.
\end{theorem}
{\bf Proof.}
According to proposition 3, 
a general Hamiltonian $(n-1)$-form $H$ 
can be written as
\begin{equation*}
  \begin{split}
    H(x,v,p)
    &={\scriptstyle\frac1{(n-1)!}} 
    \left(p^\mu_A f^A(x,v)+g^\mu(x,v)\right)
    \epsilon_{\mu\nu_1\cdots\nu_{n-1}} 
    dx^{\nu_1}\wedge\cdots\wedge dx^{\nu_{n-1}},
  \end{split}
\end{equation*}
where $f^A,g^\mu$ are pulled back functions from $\mathcal E$.
Thus the corresponding Hamiltonian vector $X$ field takes the form
\begin{equation*}
  X_{(x,v,p)}
  =
  f^A(x,v)\partial_A
  -(p^\mu_A\partial_Bf^A(x,v)+\partial_B g^\mu)\partial_\mu^B,
\end{equation*}
so $X$ is projectable as stated.\\
Let $X$ be a vertical vector field on
$\mathcal E$. $X$ generates a one-parameter group of fibre diffeomorphisms on
$\mathcal E$ which by jet prolongation can be transported to
$\mathfrak J^1\mathcal E$ (see \cite{GoIsMa98}, sec. 4B). 
As $(\mathfrak J^1\mathcal E)^\star$ is 
affine dual to the first jet bundle, one can lift this action further
by pull back and -- using the canonical projection $\pi_{\mathcal P}$
-- transport it to $\mathcal P$. For a given vector field
$X=X^A\partial_A$ the coordinate expression of the latter is as follows
\begin{equation}\label{X-hat}
  \hat X_{(x,v,p)}=X^A_{(x,v)}\partial_A
  -p^\nu_B \,\partial_A X^B_{(x,v)}\,\partial_\nu^A.
\end{equation}
One immediately verifies that such an $\hat X$ is Hamiltonian with
Hamiltonian $(n-1)$-form $I(X)$:
\begin{equation*}
  \begin{split}
    \PCPdV I(X) 
    &={\scriptstyle \frac1{(n-1)!}}\epsilon_{\mu\nu_1\cdots\nu_{n-1}}
    \left(
      \partial_BX^A \mathfrak e^B+X^A\mathfrak e^\mu_A
    \right)
    dx^{\mu_1}\wedge\cdots\wedge dx^{\mu_{n-1}}
    =
    \hat X\PCPcontraction\PCPoV.
  \end{split}
\end{equation*}
\PCPqed
Further, as the lifting of two such vector fields $X,Y$ is a
Lie-homomorphism,
$\widehat{[X,Y]}=[\hat X,\hat Y]$, and
$\hat X\PCPcontraction \hat Y\PCPcontraction\PCPoV
  =[\hat X,\hat Y]\PCPcontraction\PCPtV$,
one has
\begin{equation*}
  \{I(X),I(Y)\}=\hat X\PCPcontraction \hat Y\PCPcontraction\PCPoV
  =[\hat X,\hat Y]\PCPcontraction\PCPtV
  =I([X,Y]).
\end{equation*}
Hence the map $I$ provides a momentum map for vertical symmetries of 
$\mathcal E$.

One might wonder what the different ways to lift vertical
automorphisms to the jet bundle and to its affine dual have to do with
each other (cf. \ref{X-hat} to \ref{prol-variation}). 
The link is provided by the Legendre transformation%
: If an automorphism leaves the Lagrange density
invariant, then its lift to the first jet bundle is mapped to its
lift to the co-jet bundle via the Legendre transformation.

\begin{appendix}


\section{\label{conn-facts}Some facts about connections \cite{KMS93}}

Most generally, a connection on a fibre bundle 
$(\mathcal E,\pi,\mathcal M)$ is a projection onto the vertical
subbundle $\mathfrak V\mathcal E$.
One can show that such projections are in $1-1$ correspondence with 
maps $\Gamma:\mathcal E\rightarrow\mathfrak J^1\mathcal E$.
In coordinates, one writes
\begin{equation*}
  \Gamma:\mathcal E\ni (x^i,v^A)
  \mapsto
  (x^i,v^A,\Gamma^A_i(x,v))\in\mathfrak J^1\mathcal E.
\end{equation*}
For $(\mathcal E,\pi,\mathcal M)$ being a vector bundle and $\Gamma$
a linear connection (cf. \cite{KMS93}), one has in addition
$  \Gamma^A_i(x,v)=-\Gamma^A_{iB}(x)\,v^B$,
where $\Gamma^A_{iB}(x)$ denote the Christoffel symbols of the
corresponding covariant derivative.

\section{A connection $\bar\Gamma$ on $\mathcal P$}

If we want to construct a connection on $\mathcal P$, then we are
looking for a $(\pi^1_0)^\star$-fibre preserving map
$\bar\Gamma:\mathcal P\rightarrow\mathfrak J^1\mathcal P$,
where $\mathfrak J^1\mathcal P$ is defined w.r.t. the projection 
$(\pi_1)^\star$ onto $\mathcal M$.
For such a map, one needs both a connection $\Gamma$ on $\mathcal E$
and a connection $\Lambda$ on $T\mathcal M$ (cf. \cite{Ko87}). 
 The coordinate
expression of the desired map $\bar \Gamma$ turns out to be (cf. \cite{Pa00})
\begin{equation}\label{Gamma-bar-coord}
  \bar\Gamma:(x^\mu,v^A,p^\mu_A)\mapsto
  (x^\mu,v^A,p^\mu_A,\Gamma_\mu^A,
  (\partial_A\Gamma^B_\mu)\delta^\nu_\sigma p^\sigma_B
  +\Lambda_{\mu\sigma}^\nu\delta^B_A p^\sigma_B
  -\Lambda_{\mu\rho}^\rho\delta^B_A\delta^\nu_\sigma p^\sigma_B).
\end{equation}
Since $T\mathcal M$ is a vector bundle we can use the
Christoffel symbols $\Lambda_{\mu\nu}^\rho$ of $\Lambda$.
Moreover,
\begin{equation}\label{Gamma-spur}
  \begin{split}
    (\partial_A\Gamma^B_\mu)\delta^\mu_\sigma
    +\Lambda_{\mu\sigma}^\mu\delta^B_A 
    -\Lambda_{\mu\rho}^\rho\delta^B_A\delta^\mu_\sigma
    &=
    (\partial_A\Gamma^B_\sigma)
    +(\Lambda_{\mu\sigma}^\mu
      -\Lambda_{\sigma\mu}^\mu)\delta^B_A, 
  \end{split}
\end{equation}
but the last term vanishes for torsion free connections $\Lambda$.
With the help of this connection, we can define the a basis on
$T^\ast\mathcal P$ at every point $(x,v,p)$ by
\begin{equation}\label{Gamma-basis}
  dx^\mu, \quad \mathfrak e^A= dv^A-\Gamma_\mu^A(x,v) dx^\mu,
  \quad \mathfrak e^\mu_A=
  dp^\mu_A-\bar\Gamma_{\nu A}^\mu(x,v,p) dx^\nu.
\end{equation}
\end{appendix}


\end{document}